\documentclass[preprint,pra,aps,showpacs,preprintnumbers,amsmath,amssymb]{revtex4}

\def\v{{ \frac {V_0} {\hbar}}}

\def\ih{{ \frac{i}{\hbar} }}

\def\half{\frac{1}{2}}

\newcommand\beq{\begin{equation}}
\newcommand\eeq{\end{equation}}
\newcommand\bea{\begin{eqnarray}}
\newcommand\eea {\end{eqnarray}}

\begin{document}


\title{Path Integral Analysis of Arrival Times with a Complex
Potential}

\author{J.J.Halliwell}%

\affiliation{Blackett Laboratory \\ Imperial College \\ London SW7
2BZ \\ UK }

\date{\today}

\begin{abstract}
A number of approaches to the arrival time problem employ a
complex potential of a simple step function type and the arrival
time distribution may then be calculated using the stationary
scattering wave functions.
Here, it is shown that in the Zeno limit (in which the potential becomes
very large), the arrival time distribution may be obtained in a clear
and simple way using a path integral representation of the
propagator together with the path decomposition expansion
(in which the propagator is factored across a surface of constant time).
This method also shows that the same result is obtained for a wide
class of complex potentials.
\end{abstract}

\maketitle

\section{Introduction}

Some of the interesting outstanding problems in quantum theory
concern situations in which time appears in a non-trivial way. One
of the simplest such problems is the arrival time problem. In the
one-dimensional version of this problem one considers an initial
wave function concentrated in the region $x>0$ and consisting
entirely of negative momenta. The question is then to find the
probability that the particle crosses $x=0$ between time $ \tau$
and $ \tau + d \tau$.  The classical analysis of this problem is
trivial, but the quantum analysis is not so, due in part to the
fact that the usual machinery of quantum measurement refers to
fixed moments of time and not to measurements distributed over an
interval of time.

There are many different approaches to this problem \cite{time}. One
particular approach is to include a complex potential
\beq
V(x) =
- i V_0 \theta (-x)
\label{1}
\eeq
in the Schr\"odinger equations, and to then compute
the final state
\beq | \psi (\tau) \rangle = \exp \left( - \ih H_0
\tau - \frac { V_0} {\hbar} \theta (-x) \tau \right) | \psi \rangle
\label{2}
\eeq
where
$H_0$ is the free Hamiltonian. The intuitive idea behind this is
that the part of the wave packet that reaches the origin during
the time interval $[0, \tau]$ is absorbed, so that
\beq N( \tau )
= \langle \psi (\tau) | \psi (\tau) \rangle
\label{3}
\eeq is the
probability of not crossing during the time interval. The
probability of crossing between $ \tau$ and $\tau + d \tau $ is
then
\beq
\Pi (\tau) = - \frac { d N } { d \tau}
\label{4}
\eeq
Such potentials were originally considered by Allcock in his seminal
work \cite{All} and have subsequently appeared in detector models of
arrival times \cite{Hal1,Muga}. A recent interesting result of Echanobe et al.
is that under certain conditions a complex potential of the form
Eq.(\ref{1}) is essentially the same as pulsed measurements, in which the
wave function is measured at discrete time intervals \cite{Ech}.

The difficulty behind this approach, however, is that the wave function is not
entirely absorbed by this complex potential and $N (\tau)$ includes parts of the
initial wave function that have reflected off the potential. The reflection
is small for small $V_0$ but then the resolution of the measurement, which
is proportional to $\hbar /V_0$, is poor. On the other hand, there is a lot of
reflection for large $V_0$ and indeed the wave function is entirely reflected
in the limit $V_0 \rightarrow \infty$. This is the quantum Zeno effect \cite{Zeno}
and plagues many different approaches to the arrival
time problem.

Echanobe et al. have made an interesting proposal which embraces the Zeno effect
in the $V_0 \rightarrow \infty$ limit yet at the same time extracts the physics
hidden within it by suitable normalization \cite{Ech}. They consider the limit
$V_0 \rightarrow \infty$ of the expression
\beq
\Pi_N(\tau) = \frac{ \Pi (\tau)  } { 1 - N (\infty) }
\label{5}
\eeq
which is normalized since
$
\int_0^{\infty} d \tau \ \Pi (\tau ) = 1 - N( \infty).
$
The point is that $N (\infty)$ represents the total amount of reflected
wave function, so $ 1 - N (\infty)$ represents the total probability
of crossing during the time interval $ [ 0, \infty ) $ and this goes
to zero  as $V_0 \rightarrow \infty$. The expression $ \Pi ( \tau ) $
also goes to zero as $ V_0 \rightarrow \infty$ but the ratio Eq.(\ref{5})
is finite and independent of $V_0$, and defines a reasonable normalized
arrival time distribution function.
Known results on the
stationary scattering wave functions \cite{Muga2} yield the result
\beq
\Pi (\tau) =\frac { 2} { m^{3/2} V_0^{1/2}}
\langle \psi_f (\tau) |\hat p \delta ( \hat x ) \hat p
| \psi_f (\tau) \rangle
\label{6}
\eeq
from which the normalized result is
\beq
\Pi_N (\tau) = \frac { \hbar } { m \langle p \rangle }
\langle \psi_f (\tau) |\hat p \delta ( \hat x ) \hat p
| \psi_f (\tau) \rangle
\label{7}
\eeq
where $ | \psi_f (\tau) \rangle $ is the freely evolved wave function and
$\langle p \rangle $ is the average momentum in the initial wave packet \cite{Ech,Muga2}.

An interesting question in these expressions is the origin of the form of
the $ \hat p \delta (\hat x) \hat p $ term, which is not obvious from the
calculation of it in Ref.\cite{Muga2}.
Compare this to the simplest guess for the arrival
time distribution function, the current density
\beq
J(t) = \frac { \hbar } {2 m }
\langle \psi_f (\tau) | \left( \hat p \delta ( \hat x ) + \delta ( \hat x ) \hat p \right)
| \psi_f (\tau) \rangle
\label{8}
\eeq
which is sensible classically but is not always positive in the quantum case \cite{cur}.
(Here $ \delta (\hat x ) = | 0 \rangle \langle 0 | $, where $ | 0 \rangle $ denotes the
eigenstate $ | x \rangle $ of the position operator at $x=0$.)
A simple operator re-ordering of $J(t)$ gives the ``ideal" arrival time distribution
of Kijowski
\beq
\Pi_K (\tau) = \frac { \hbar } { m  }
\langle \psi_f (\tau) |\hat p^{1/2} \delta ( \hat x ) \hat p^{1/2}
| \psi_f (\tau) \rangle
\label{9}
\eeq
which is clearly positive, but harder to relate to specific measurement schemes
\cite{Kij}. It would be of value to find a derivation of Eq.(\ref{7}) which gives
some insight into the form of this expression.
A further question concerns the role of the complex potential. Although Eq.(\ref{7})
is independent of $V_0$, it is not clear to what extent the result depends
on the specific choice of complex potential, Eq.(\ref{1}).

In this paper, we show that both of these questions are answered very simply
using path integral methods. We show that the general form (\ref{6}) is derived very
easily (up to a constant, fixed by normalization). Secondly, the result (\ref{7})
is seen to be true for a wide class of
potentials of the form
\beq
V(x) = - i V_0 \theta (-x) f(x)
\label{10}
\eeq
where $f(x)$ is any positive function.

With the general complex potential Eq.(\ref{10}), the arrival time distribution
(\ref{4}) is given by
\bea
\Pi (\tau) &=& 2  \langle \psi_\tau | V( \hat x) | \psi_\tau \rangle
\nonumber \\
&=& 2 \frac {V_0} {\hbar} \int_{-\infty}^0 dx \ f (x)  \left| \psi (x,\tau) \right|^2
\label{11}
\eea
where $\psi (x,\tau)$ is defined in Eq.(\ref{2}).
The key is therefore to evaluate the propagator
\beq
g(x_1, \tau | x_0 ,0 ) = \langle x_1 | \exp \left( - \ih H_0 \tau - \v_0 \theta (- \hat x) f(\hat x) \tau \right)
| x_0 \rangle
\label{12}
\eeq
for $x_1 < 0 $ and $ x_0 > 0$. This may be calculated using a sum over paths,
\beq
g(x_1, \tau | x_0,0 ) = \int {\cal D} x \exp \left( \ih S \right)
\label{13}
\eeq
where
\beq
S = \int_0^{\tau} dt \left( \half m \dot x^2 + i V_0 \theta (-x) f(x) \right)
\label{14}
\eeq
and the sum is over all paths $ x(t)$ from $x(0) = x_0$ to $x(\tau) = x_1$.

To deal with the step function form of the potential we need to split off the sections
of the paths lying entirely in $x>0$ or $x<0$. The way to do this is to use
the path decomposition expansion
or PDX \cite{PDX,HaOr,Hal2}.
Each path from $x_0>0$ to $x_1 <0$ will typically cross $x=0$
many times, but all paths have a first crossing, at time $t_1$, say. As a consequence
of this, it is possible to derive the formula,
\beq
g(x_1, \tau | x_0,0 ) = \frac {i \hbar } {2m} \int_{0}^{\tau} dt_1
\ g (x_1, \tau | 0, t_1) \frac {\partial g_r } { \partial x} (x,t_1| x_0,0) \big|_{x=0}
\label{15}
\eeq
Here, $g_r (x,t|x_0,0)$ is the restricted propagator given by a
sum over paths of the form (\ref{13}) but with all paths restricted to
$x(t) >0$. It vanishes when either end point is the origin but its derivative
at $x=0$ is non-zero (and in fact the derivative of $g_r$ corresponds to
a sum over all paths in $x>0$ which end on $x=0$ \cite{HaOr}). It is also useful
to record a PDX formula involving the last crossing time $t_2$,
\beq
g(x_1, \tau | x_0,0 ) =  - \frac {i \hbar  } {2m} \int_{0}^{\tau} dt_2
\ \frac {\partial g_r} {\partial x} (x_1, \tau | x, t_2) \big|_{x=0} \ g (0,t_2|x_0,t_1)
\label{16}
\eeq
These two formulae may be combined to give a first and last crossing version of the PDX,
\beq
g(x_1, \tau | x_0,0 ) =  \frac {\hbar^2 } {4m^2} \int_{0}^{\tau} dt_2
\int_0^{t_2} dt_1
\ \frac {\partial g_r} {\partial x} (x_1, \tau | x, t_2) \big|_{x=0} \ g (0,t_2| 0,t_1)
\ \frac {\partial g_r } { \partial x} (x,t_1| x_0,0) \big|_{x=0}
\label{16a}
\eeq
The restricted propagators in the two regions are easier to work with
than Eq.(\ref{12}) since
the potential is zero throughout $x>0$ and $ - i V_0 f(x)$ throughout $x<0$
so these may be calculated by standard methods, without the complication of
the $ \theta (-x)$ term.
The problem of calculating the propagator
Eq.(\ref{12}) therefore essentially reduces to the easier problem
of calculating it between two points lying on $x=0$. This can sometimes
be evaluated by a mode sum calculation \cite{Car}, even though the full propagator
Eq.(\ref{12}) is not necessarily calculable in this way.

Returning to the first crossing PDX, Eq.(\ref{15}),  $g_r$
is the restricted propagator for the {\it free particle},
which is given by the method of images expression
\beq
g_r (x_1, \tau |x_0,0) = \theta (x_1 ) \theta (x_0)
\left( g_f (x_1, \tau |x_0,0) - g_f (-x_1, \tau |x_0,0) \right)
\label{17}
\eeq
where $g_f $ denotes the free particle propagator.
It follows that
\beq
\frac {\partial g_r } { \partial x} (x,t_1| x_0,0) \big|_{x=0} =
2 \frac {\partial g } { \partial x} (0,t_1| x_0,0) \theta (x_0)
\label{18}
\eeq
Inserting this in Eq.(\ref{15}), and rewriting it as an operator expression, we obtain
the result
\bea
\langle  x_1 |   & \exp & \left( - \ih H_0 \tau  - \v \theta (-\hat x) f(\hat x) \tau  \right) | x_0 \rangle
\nonumber \\
=  &-&  \frac {1} {m} \int_0^{\tau} dt \ \langle x_1 | \exp \left( - \ih H_0 (\tau -t) - \v \theta (-\hat x)
f(\hat x)  (\tau -t) \right)
\nonumber \\
& \times & \delta  ( \hat x ) \hat p \ \exp \left( - \ih H_0 t \right)  | x_0 \rangle
\label{19}
\eea
We note the appearance of the combination $\delta (\hat x) \hat p $ which clearly
corresponds to a ``crossing operator" (and in derivations of the PDX arises directly
from a derivative of $ \theta (\hat x)$ \cite{Hal2}).
Now note that the operator $\delta (\hat x)$ has the simple
property that for any operator $A$
\beq
\delta (\hat x) A \delta (\hat x) = \delta ( \hat x ) \langle 0 | A | 0 \rangle
\label{20}
\eeq
This property together with Eq.(\ref{19}) inserted in Eq.(\ref{11}) yields
\bea
\Pi (\tau) &=& \frac {2 V_0 } {\hbar m^2}
\int_0^\tau dt' \int_0^{\tau} dt \int_{-\infty}^0 dx \ f(x)
\nonumber \\
& \times & \langle 0 | \exp \left(  \ih H^{\dag}  (\tau - t')\right) | x \rangle \langle x | \exp \left(
- \ih H (\tau - t)\right) | 0 \rangle
\nonumber \\
& \times & \langle \psi | \exp \left( \ih H_0 t' \right)  \hat p \ \delta (\hat x) \ \hat p \ \exp \left(
 - \ih H_0 t \right) | \psi \rangle
\label{21}
\eea
where $H = H_0 - i V_0 \theta (-x) f(x) $ is the total (non-hermitian) Hamiltonian. Now the key point is that
as $V_0 \rightarrow \infty$, the integrals over $t$ and $t'$ are strongly
concentrated around $\tau$, so we obtain
\beq
\Pi (\tau) \approx C \ \langle \psi | \exp \left( \ih H_0 \tau \right)  \hat p \ \delta (\hat x) \ \hat p \ \exp \left(
 - \ih H_0 \tau \right) | \psi \rangle
\label{22}
\eeq
where
\bea
C &=& \frac {2 V_0 } {\hbar m^2}
\int_0^\tau dt' \int_0^{\tau} dt \int_{-\infty}^0 dx \ f(x)
\nonumber \\
& \times & \langle 0 | \exp \left(  \ih H^{\dag}  (\tau - t')\right) | x \rangle \langle x | \exp \left(
- \ih H (\tau - t)\right) | 0 \rangle
\label{23}
\eea
Furthermore, by the change of variables $ s= \tau - t $, $ s' = \tau - t'$,
it is easily seen that $C$ is independent of $\tau$ for large $V_0$, so is just
a constant. It is also easily seen that the approximation (\ref{22}) will hold
for $V_0 \gg k^2 / 2 m $, where $|k|$ is the largest momentum in the initial state.

Eq.(\ref{22}) is the main result and is of precisely the form Eq.(\ref{6}), up to an overall
constant. Since Eq.(\ref{7}) is obtained by normalization of Eq.(\ref{6}), we
therefore obtain Eq.(\ref{7}). It holds for a general
class of potentials, of the form Eq.(\ref{10}), not just the simple case
$ f(x) = 1$ considered in Refs.\cite{Ech,Muga2}. The reason for this is that
for large $V_0$ the paths in the path integral are kept out of the region
$x<0$, only entering it at just before the final time, so the result has
very limited dependence on the detailed form of the potential in $x<0$.
The result Eq.(\ref{22}), and in particular
the $ \hat p \delta ( \hat x ) \hat p $ form,
follow as an almost immediate consequence
of the PDX for large $V_0$,
together with the simple property Eq.(\ref{20}).

To fully verify the validity of this path integral approach,
we now calculate the constant $C$ in the case $ f(x) = 1$
and look for detailed agreement with the unnormalized result Eq.(\ref{6}).
Eq.(\ref{23}) may be written,
\beq
C = \frac {2 V_0} { \hbar m^2} \int_{-\infty}^0 dx \ | \phi (x) |^2
\label{24}
\eeq
where, after a change of variables $s = \tau - t$,
\beq
\phi (x) = \int_0^{\tau} ds \ \langle x |
\exp \left( - \ih H_0 s  - \v \theta (-\hat x)  s  \right) | 0 \rangle
\label{25}
\eeq
The integrand may be represented as a sum over paths from $x=0$ at time
$s=0$ to the final point $x<0$ at time $\tau$. We may use the last crossing
PDX, Eq.(\ref{16}), to split this into a sum over paths from $x=0$ at $s=0$,
to the last crossing at $x=0$, $s=s_1$, and from there propagating entirely
in $x<0$ to the final $x$ at time $s=\tau$. This reads
\beq
\phi (x) = \frac {1} {m} \int_0^{\tau} ds \ \int_0^s du \ \langle x |
\exp \left( - \ih H_0 (s-u)  - \v   (s-u)  \right) \hat p | 0 \rangle
\langle 0 | \exp \left( - \ih H u \right) | 0 \rangle
\label{26}
\eeq
In the second bracket expression, $H$ is the total Hamiltonian so
still includes the $\theta (-x)$ potential. This is the propagator
along the edge of an imaginary step potential which, fortunately
was calculated in Ref.\cite{Car} using a mode sum method
(for the case of a real step potential,
but this is readily continued to imaginary values), and we have
\beq
\langle 0 | \exp \left( - \ih H u \right) | 0 \rangle
= \left( \frac {m} {2 \pi i \hbar } \right)^{1/2}
\ \frac { (1 - e^{- V_0 u / \hbar }) } { (V_0/\hbar) u^{3/2}}
\label{27}
\eeq
Noting that
\beq
\int_0^{\tau} ds \ \int_0^s du = \int_0^{\tau} du \ \int_u^\tau ds
\label{28}
\eeq
and changing variables to $ v = s - u$, we see that for large $V_0$
the dominant contribution comes from close to $u=0$ and $v=0$.
We may therefore let $\tau \rightarrow \infty$, and the integrals
then factor into a product
\beq
\phi (x) = \frac {1} {m} \ \int_0^{\infty} dv
\ \langle x |
\exp \left( - \ih H_0 v  - \v  v  \right) \hat p | 0 \rangle
\ \int_0^{\infty} du \ \left( \frac {m} {2 \pi i \hbar } \right)^{1/2}
\ \frac { (1 - e^{- V_0 u / \hbar }) } { (V_0/\hbar) u^{3/2}}
\label{29}
\eeq
The first integral may be evaluated using the familiar formula \cite{Sch}
\beq
\int_0^{\infty} dt \ \left( \frac {m} {2 \pi i \hbar t } \right)^{1/2}
\exp \left( \ih \left[ Et + \frac { m x^2} {2 t} \right] \right)
= \left( \frac { m } { 2 E} \right)^{\half} \ \exp \left( \ih | x | \sqrt {2 m E} \right)
\label{30}
\eeq
with $E = i V_0 $, and then applying $ - i \hbar \partial / \partial x $.
The second is evaluated using the formula,
\beq
\int_0^{\infty} dx \ \frac {( 1 - e^{-x} )} {x^{3/2} } = 2 \sqrt {\pi}
\label{31}
\eeq
We thus obtain
\beq
\phi (x) = \left( \frac { 2 m} {V_0} \right)^{1/2} \exp \left(
- \frac {(1-i)} {\hbar}  \sqrt{ m V_0} | x  | \right)
\label{32}
\eeq
Inserting in Eq.(\ref{24}), this gives the final result
\beq
C = \frac {2 } { m^{3/2} V_0^{1/2} }
\label{33}
\eeq
The path integral method used here therefore gives
precise agreement with the earlier result Eq.(\ref{6})
obtained by scattering methods.

\bibliography{apssamp}

\end{document}